\documentclass[aps,prd,nofootinbib,groupedaddress,twocolumn,floatfix]{revtex4}
\usepackage{graphicx}
\usepackage{epsfig}
\usepackage{dcolumn}
\usepackage{amsmath}
\def\Journal#1#2#3#4{{#1} {\bf#2}, #3 (#4)}

\def\NPA{{\rm Nucl. Phys.} A}
\def\NPB{{\rm Nucl. Phys.} B}
\def\PLB{{\rm Phys. Lett.}  B}
\def\PRL{\rm Phys. Rev. Lett.}
\def\PRD{{\rm Phys. Rev.} D}
\def\PRC{{\rm Phys. Rev.} C}

\def\EPJC{{\rm Eur. Phys. J.}C}

\def\la{\langle}
\def\ra{\rangle}

\def\be{\begin{equation}}
\def\ee{\end{equation}}
\def\bea{\begin{eqnarray}}
\def\eea{\end{eqnarray}}
\input{epsfig.sty}
\begin{document}
\title{Conformal Symmetry and Pion Form Factor: Soft and Hard 
Contributions}
\author{ Ho-Meoyng Choi$^{a}$ and Chueng-Ryong Ji$^{b}$\\
$^a$ Department of Physics, Teachers College, Kyungpook National University,
     Daegu, Korea 702-701\\
$^b$ Department of Physics, North Carolina State University,
Raleigh, NC 27695-8202}
\begin{abstract}
We discuss a constraint of conformal symmetry in the analysis of the pion form 
factor. The usual power-law behavior of the form factor obtained in the
perturbative QCD analysis can also be attained by taking negligible quark 
masses in the nonperturbative quark model analysis, confirming the recent 
AdS/CFT correspondence. 
We analyze the transition from soft to hard contributions in 
the pion form factor considering a momentum-dependent dynamical quark mass from 
an appreciable constituent quark mass at low momentum region 
to a negligible current quark mass at high momentum region. 
We find a correlation between the shape of nonperturbative quark distribution
amplitude and the amount of soft and hard contributions to the pion form factor.  
\end{abstract}


\maketitle

\section{Introduction}
The pion elastic form factor has been exemplified by many calculations
to understand the substructure of hadrons in terms of the quark-gluon degrees 
of freedom~\cite{BL,IL,Ra,LS,Jacob,JK,ADT,SSK,HW,JPS,SJR}.
Since the valence structure of the pion is relatively
simple, the value of the four-momentum transfer
square $Q^2$ above which a perturbative QCD(PQCD) approach 
can be applied to the
pion structure is expected to be lower than the case of the nucleon.

Recent discussions on the anti-de Sitter space geometry/conformal field 
theory(AdS/CFT) correspondence reveals a remarkable
consistency with the QCD predictions on both hadron mass spectra and
electromagnetic form factors~\cite{BT,Ra_AdS}. 
In particular, the power-law behavior of
the pion form factor $F(Q^2) \sim 1/Q^2$ is well reproduced by the
AdS/CFT correspondence. The key ingredient in this correspondence
is the conformal symmetry valid in the negligible quark masses.
In this work, we confirm that the power-law behavior of the pion form 
factor is indeed attained by taking into account a momentum dependent dynamical
quark mass which becomes negligibly small at large momentum region
even in the nonperturbative quark model analysis.
This result is consistent with an important point of the AdS/CFT prediction,
namely the holographic wavefunction contains the contribution from
all scales up to the confining scale. Thus, it is rather unnatural to 
distinguish
the soft and hard pion form factors in comparing the predictions
between QCD and AdS/CFT. However, due to the ongoing debate on the PQCD 
applicability in exclusive processes,
it is of special interest in the QCD side 
to study the transition from the soft regime
governed by all kinds of quark-gluon correlations at low $Q^2$ to the
perturbative regime at high $Q^2$. 
We thus discuss a correlation between
the shape of nonperturbative quark distribution amplitude(DA) and the amount
of soft and hard contributions to the pion form factor utilizing our light-front
quark model (LFQM)\cite{CJ99}. Similar to the previous findings from the
Sudakov suppression\cite{LS,Jacob,JK,ADT} of the soft contribution,
we note that the soft and hard contributions are correlated with each other,
i.e. the suppression of the endpoint region for the quark DA 
corresponds to the suppression of the soft contribution, or equivalently,
the enhancement of the hard contribution. 
We further confirm that the higher helicity components suppress the 
contributions from the ordinary helicity components.
  
The paper is organized as follows. In the next section (Section II),
we present the soft contribution to the pion form factor using the 
LFQM and discuss a consistency with the AdS/CFT correspondence.
In section III, we discuss the hard contribution to the pion form factor 
including the intrinsic transverse momentum effect. 
In section IV, we show our numerical results 
of the nonperturbative quark model predictions on the pion DA and 
form factor as well as the PQCD prediction on the pion form factor. 
The correlation between soft and hard contributions and the 
higher helicity contribution are also
discussed in this section. Summary and conclusion follow in Section V.

\section{Pion Form Factor in LFQM}

To discuss nonperturbative quark model prediction of the pion form factor,
we briefly summarize our LFQM\cite{CJ99} first.
The model wave function for a pion~\cite{CJ99} is given by
\bea\label{w.f}
\Psi(x,{\bf k}_\perp,\lambda\bar{\lambda})
= \phi_R(x,{\bf k}_\perp) {\cal R}_{\lambda\bar{\lambda}}(x,{\bf k}_\perp),
\eea
where $\phi_R(x,{\bf k}_\perp)$ is the radial wave function and 
${\cal R}_{\lambda\bar{\lambda}}(x,{\bf k}_\perp)$ is the spin-orbit
wave function obtained by the interaction-independent Melosh 
transformation~\cite{Mel}
from the ordinary equal-time static spin-orbit wave function assigned by
the quantum numbers $J^{PC}=0^{-+}$. Here, $\lambda+\bar\lambda=0$ and 1 
represent the contributions from the ordinary-helicity and the higher-helicity
components, respectively. 
The detailed description for the spin-orbit wave function
can be found in~\cite{CJ99} as well as in other literatures~\cite{PJ}.
The pion wave function $\Psi$ is
represented by the Lorentz-invariant variables $x_i=p^+_i/P^+$, 
${\bf k}_{\perp i}={\bf p}_{\perp i}- x_i{\bf P}_\perp$ and $\lambda_i$, 
where $P,p_i$ and $\lambda_i$ are the meson momentum, the 
momenta and helicities 
of the constituent quarks, respectively.
The radial wave function is given by
\bea\label{radial}
\phi_R(x,{\bf k}_\perp)= \sqrt{\frac{1}{\pi^{3/2}\beta^3}}
\exp(-{\vec k}^2/2\beta^2),
\eea
where the gaussian parameter $\beta$ is related with the size of pion and
the three momentum squared ${\vec k}^2$ in terms of the light-front (LF) 
variables is given by 
\bea\label{ksquare}
{\vec k}^2 = \frac{{\bf k}_\perp^2 + m^2}{4x(1-x)}-m^2,
\eea
for the quark mass $m_u = m_d = m$.
Here, it is easy to see that the invariant mass of the pion
$M_0 = \sqrt{ \frac{{\bf k}_\perp^2 + m^2}{x(1-x)}} = 
2\sqrt{{\vec k}^2 + m^2}$ and 
the longitudinal component $k_z$ of the three momentum is given
by $k_z=(x-1/2)M_0$.

For the low momentum transfer phenomenology in LFQM, it is customary
to take a constant constituent quark mass $m$ as a mean value of 
the momentum dependent dynamical quark mass at low momentum region.
The momentum dependence of the dynamical quark mass in the spacelike
momentum region has been discussed in lattice QCD~\cite{Wil} as well as 
in other approaches, especially in 
Dyson-Schwinger~\cite{DS1,DS2} approach. Although the exact form of $m({\vec k}^2)$
is still not known, the evolution of $m({\vec k}^2)$ to a small current
quark mass in large ${\vec k}^2$ region is generally agreed.
Thus, for the large momentum transfer phenomenology in PQCD, it seems 
reasonable to take a negligible current quark mass, $m \approx 0$, for the
pion form factor.

Matching between the low momentum LFQM prediction and the large momentum
PQCD prediction is a highly nontrivial task which goes beyond the scope
of our present work. Nevertheless, a sort of consistency between LFQM and PQCD
predictions can be achieved by taking into account a momentum dependent
dynamical quark mass in the following sense. 

The PQCD factorization of the pion form factor $F_\pi (Q^2)$
can be found by taking a large $Q^2$ limit of the LFQM formula,
i.e. the convolution of the initial and final LFQM wave functions.
Since the LFQM wave function satisfies a LF bound-state equation, one can iterate
the final wave function at the momentum scale $Q$ to yield an irreducible 
scattering kernel convoluted with a quark DA which is a 
transverse-momentum-integrated wave function collinear up to the scale $Q$.
Similarly, the factorized initial quark DA is also at the scale $Q$ which is
the probing momentum scale of the virtual photon. Thus, the factorized
PQCD amplitude of $F_\pi (Q^2)$ is symmetric under the exchange of the initial
and final quark DAs. This implies that the effective quark degrees of freedom
probed by the virtual photon in computing the main contribution to $F_\pi (Q^2)$
has the momentum scale $Q$. 

From this picture, one may parametrize the probed quark mass scale as 
the momentum scale $Q$ of the virtual photon and consider the effective
dynamical quark mass as $m(Q^2)$. For the low and high $Q^2$, $m(Q^2)$ 
corresponds to the consituent and current quark mass, respectively.
Thus, one may treat the quark mass in LFQM for the low $Q^2$ phenomenology 
as a consituent quark mass and the quark mass in PQCD for the high $Q^2$ 
phenomenology as a current quark mass. For simplicity in discussing the main
features of soft and hard form factor, one may take each 
(consituent or current) mass as a constant mean value of 
$m(Q^2)$ in each  (low or high) $Q$ region.

In our calculation of both soft and hard form factor, 
we adopt the Drell-Yan-West
frame($q^+$=$q^0+q^3$=0) with ${\bf q}^2_\perp$=$Q^2$=$-q^2$. The momentum
assignment in this $q^+=0$ frame for the pion form factor is given by 
$P=(P^+, M^2_\pi/P^+,0)$ and $q=(0,{\bf q}^2_\perp/P^+,{\bf q}_\perp)$.
In this frame, the charge form factor of the pion, 
$\la P+q|J^\mu|P\ra = (2P +q)^\mu F_\pi (Q^2)$, 
can be expressed for the 
``+"-component of the current $J^\mu$ as follows~\cite{CJ99}:
\bea\label{soft_ff}
F^{\rm LFQM}_{\pi}\hspace{-0.2cm}&=&\hspace{-0.2cm}
\int^1_0 \hspace{-0.2cm}dx\int d^2{\bf k}_\perp
\sqrt{\frac{\partial k_z}{\partial x}}\phi_R(x,{\bf k}_\perp)
\sqrt{\frac{\partial k'_z}{\partial x}}\phi_R(x,{\bf k'}_\perp)
\nonumber\\
&&\times
\frac{m^2 + {\bf k}_\perp\cdot{\bf k'}_\perp}
{\sqrt{m^2+{\bf k}^2_\perp}\sqrt{m^2+{\bf k'}^2_\perp}},
\eea
where ${\bf k'}_\perp = {\bf k}_\perp + (1-x){\bf q}_\perp$ and the 
factors $m^2$ and ${\bf k}_\perp\cdot{\bf k'}_\perp$ in the numerator come 
from the ordinary helicity($\lambda+\bar\lambda=0$) and the higher 
helicity($\lambda+\bar\lambda=\pm 1$) components of the spin-orbit wave 
function ${\cal R}_{\lambda\bar\lambda}(x,{\bf k}_\perp)$, respectively.
As discussed above, one should understand $m$ as a function of $Q^2$ in 
principle  although in practice $m(Q^2)$ for the low $Q^2$ phenomenology 
can be taken as a constant constituent quark mass. 
For the constant constituent quark 
mass $m$, our LFQM prediction of the pion form factor provides a gaussian 
fall-off at high $Q^2$ region as expected from the gaussian form of our 
radial wave function $\phi_R(x,{\bf k}_\perp)$ given by Eq.~(\ref{radial}). 
In this case, the AdS/CFT correspondence doesn't work due to the
appreciable constituent quark mass which breaks the conformal symmetry. 
However, it is remarkable that $F_\pi(Q^2)$ in Eq.~(\ref{soft_ff}) reveals 
a power-law behavior even with our gaussian radial wave function 
$\phi_R(x,{\bf k}_\perp)$ if $m$ is replaced by $m(Q^2)$ yielding a negligible 
current quark mass at large $Q^2$.  
We note that $\phi_R(x,{\bf k}_\perp) \phi_R(x,{\bf k'}_\perp)$ in 
Eq.(\ref{soft_ff}) provides a mass-dependent weighting factor 
$e^{-\frac{m^2}{4x(1-x)\beta^2}}$
which severely suppresses the contribution from the endpoint region of
$x \rightarrow 0$ and 1 unless $m \rightarrow 0$.
When the conformal symmetry limit ($m \rightarrow 0$) is taken, however,
there is no such suppression of the endpoint region and the high $Q^2$ 
behavior of the form factor dramatically changes from a gaussian fall-off to
a power-law reduction. This may partly explain why the power-law behavior 
attained in our Eq.(\ref{soft_ff}) is not accidental but a consequence of 
the constraint taken from the conformal symmetry. 
While our result $Q^n F_\pi (Q^2) \rightarrow const.$
confirms the recent findings of the AdS/CFT 
correspondence\cite{BT,Ra_AdS},
the power $n$, e.g. $n=2$ or 4, still depends on the details of the LFQM
calculation such as whether one takes into account the Jacobi factor or 
not in Eq.(\ref{soft_ff}).
The detailed numerical results are presented in section IV. 
 
\section{Hard Contribution in PQCD} 

At high momentum transfer, the pion form factor 
in leading order  
can be calculated within the PQCD from the soft part of the wave function
by means of a homogeneous Bethe-Salpeter equation. 
Absorbing the perturbative kernel of the Bethe-Salpeter equation in a hard
scattering amplitude $T_H$ containing all two-particle irreducible amplitudes
for $\gamma^* + q{\bar q}\to q{\bar q}$, 
the hard contribution to the pion electromagnetic
form factor is given by
\begin{eqnarray}\label{F_PQCD}
F^{\rm PQCD}_{\pi}(Q^2)&=&
\int \frac{d^3k d^3l}{16\pi^3}\phi_R(y,{\bf l}_\perp)
{\cal T}_H \phi_R(x,{\bf k}_\perp),
\end{eqnarray}
where 
$d^3 k= dx d^2{\bf k}_\perp\sqrt{\partial k_z/\partial x}$ 
($d^3 l= dy d^2{\bf l}_\perp\sqrt{\partial l_z/\partial y}$) and
${\cal T}_H$ is related to the original $T_H$ including  
the spin-orbit wave function, i.e.
\bea\label{TH}
{\cal T}_H = {\cal R}_0 T_H^{(\lambda+\bar{\lambda}=0)}
+ {\cal R}_{\pm 1} T_H^{(\lambda+\bar{\lambda}=\pm1)},
\eea
with
${\cal R}_{0(\pm1)}$ = ${\cal R}^*_{\uparrow\downarrow(\uparrow\uparrow)}
(y,{\bf l}_\perp) 
{\cal R}_{\uparrow\downarrow(\uparrow\uparrow)}(x,{\bf k}_\perp)$
+ ($\uparrow\leftrightarrow\downarrow$).
The hard scattering amplitudes
$T_H^{(\lambda+\bar{\lambda}=0)}$ and 
$T_H^{(\lambda+\bar{\lambda}=\pm 1)}$ in Eq.~(\ref{TH}) represent the
contributions from the ordinary-helicity and the higher-helicity components,
respectively. 

\begin{figure}[t]
\includegraphics[width=3.2in,height=2.0in]{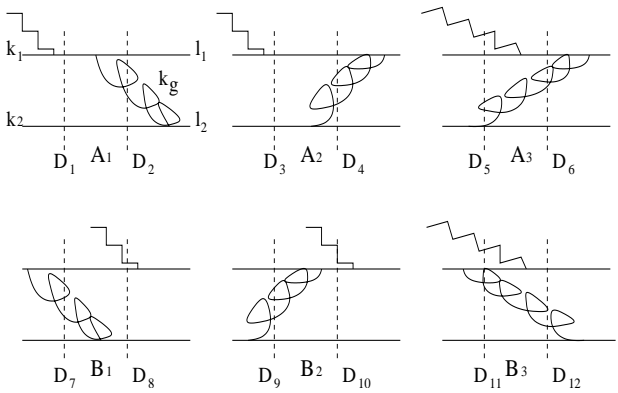}
\caption{Leading order light-front time-ordered diagrams for the pion 
form factor.}
\label{fig1}
\end{figure}

To lowest order in perturbation theory, the hard scattering amplitude 
$T_H$ is  to be calculated from the time-ordered one-gluon-exchange 
diagrams shown in Fig.~\ref{fig1}.  
The internal momenta for $(+,\perp)$-components 
are given by
$k_1= (x_1 P^+_1, {\bf k}_\perp)$,
$k_2= (x_2 P^+_1, -{\bf k}_\perp)$, 
$l_1 = (y_1P^+_1, y_1{\bf q}_\perp + {\bf l}_\perp)$, and
$l_2=(y_2 P^+_1, y_2{\bf q}_\perp - {\bf l}_\perp)$, where
$x_1=x,x_2=1-x,y_1=y$, and $y_2=1-y$.

The explicit forms for the hard scattering amplitudes including
the higher twist effects such as the quark mass $m$ and the intrinsic
transverse momenta ${\bf k}_\perp$ and ${\bf l}_\perp$
have been presented in Ref.~\cite{DD}. There seems no need 
to rewrite them in this work since the only difference in this
work is to take the quark mass as a negligibly small current quark mass, 
$m \approx 0$.
Including the intrinsic transverse momenta,
the leading contribution of the hard scattering amplitude for the higher
helicity components is of $1/Q^4$, which is the next-to-leading contribution.
When the intrinsic transverse momenta are neglected, we find by the power
counting that the higher helicity contributions goes to zero and
the hard scattering amplitude for ordinary helicity
components reduces to the usual leading twist result:
\bea\label{TH_pi}
T_H &=&\frac{16\pi \alpha_s C_F}{Q^2}
\biggl(\frac{e_u}{x_2y_2}+\frac{e_{\bar{d}}}{x_1y_1}\biggr),
\eea
where $\alpha_s=4\pi/[(11-2n_f/3){\rm Log}(Q^2/\Lambda^2)]$ 
is the QCD running coupling constant and $C_F(=4/3)$ is
the color factor. 
For the pion form factor in PQCD for large $Q^2$, 
we neglect the terms not only the quark mass $m$ but also 
${\bf k}^2_\perp/{\bf q}^2_\perp$,
${\bf l}^2_\perp/{\bf q}^2_\perp$, and
${\bf k}_\perp\cdot{\bf l}_\perp/{\bf q}^2_\perp$
both in the energy denominators and the numerators of the hard scattering
amplitude $T_H$ due to 
${\bf k}^2_\perp\ll{\bf q}^2_\perp$ and ${\bf l}^2_\perp\ll{\bf q}^2_\perp$.
Our analytic forms for the hard scattering amplitudes 
are the same as those obtained by Huang et.al~\cite{HW}. 
However, in choosing the radial wave
function as a nonperturbative input, we include the Jacobi factor 
$\sqrt{M_0/4x(1-x)}$ while the authors in~\cite{HW} do not. 
This makes quantitative difference for the soft and hard form factors between
the ones in~\cite{HW} and ours. Although the qualitative behaviors are 
equivalent to each other between the two(ours and Ref.~\cite{HW}) at high 
$Q^2$ in PQCD, we discuss the effect of the Jacobi factor with respect to 
the AdS/CFT correspondence in our numerical results.

\section{Numerical results}
The key idea in our LFQM~\cite{CJ99} is to treat the gaussian radial
wave function
as a trial function for the variational principle to the QCD-motivated
Hamiltonian saturating the Fock state expansion by the constituent quark
and antiquark, i.e. $H_{q\bar{q}}=H_0+V_{\rm int}$, where the interaction
potential $V_{\rm int}$ consists of confining and hyperfine interaction
terms.  For the numerical calculations, we 
obtained two different
sets of model parameters, i.e. (1) $(m,\beta)=(0.25, 0.3194)$ GeV for
the harmonic oscillator(HO) confining potential, and
(2) $(m,\beta)=(0.22,0.3659)$ GeV for the linear confining potential,
from our variational principle for the QCD-motivated effective
Hamiltonian~\cite{CJ99}.
Both parameter sets have shown to provide a good agreement with the available
experimental data for form factors, decay constants and charge radii etc. of
various pseudoscalar and vector mesons as well as their radiative decay
widths~\cite{CJ99}.

The quark DA of pion, $\phi(x,\mu)$,
i.e. the probability of finding collinear quarks up to the scale
$\mu$ in the $L_z=0$($s$-wave) projection of the pion
wave function~\cite{BL} is defined by
\bea\label{DA}
\phi(x,\mu)&=&\int^{{\bf k}^2_\perp<\mu^2}
\frac{d^2 {\bf k}_\perp}{\sqrt{16\pi^3}}
\sqrt{\frac{\partial k_z}{\partial x}}
\Psi(x,{\bf k}_\perp,\lambda\bar{\lambda}),
\eea
where
$\Psi(x,{\bf k}_\perp,\lambda\bar{\lambda})$=$\phi_R(x,{\bf k}_\perp)
({\cal R}_{\uparrow\downarrow} - {\cal R}_{\downarrow\uparrow})/\sqrt{2}$
and the higher helicity components($\uparrow\uparrow$ and
$\downarrow\downarrow$) do not contribute to DA in this case.
Here, again the quark mass in Eq.~(\ref{DA}) is the renormalization scale $\mu$
dependent. However, for the LFQM phenomenology at low momentum scale, one may
effectively take $m(\mu)$ as a constant constituent quark mass.

\begin{figure}
\vspace{0.7cm}
\includegraphics[height=3in,width=3in]{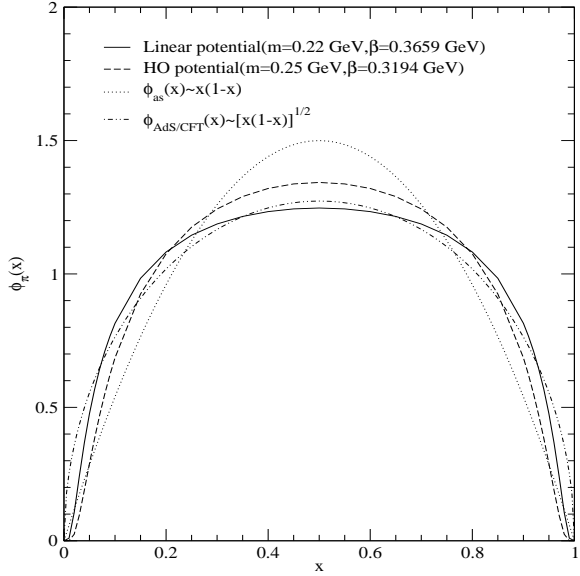}
\caption{The distribution amplitude for the pion using two different 
sets of model parameters,
(1) $(m, \beta)=(0.25, 0.3194)$ GeV(dotted line), and
(2) $(m, \beta)=(0.22, 0.3659)$ GeV(solid line) compared with the usual
QCD asymptotic DA(dotted line) as well as the AdS/CFT prediction of the
asymptotic DA(double-dot-dashed line).}
\label{fig2}
\end{figure}
In Fig.~\ref{fig2}, we show the 
normalized
quark DAs of the pion obtained 
from the linear(solid line) and HO(dashed line) potential
parameters, respectively, in comparison with the usual
QCD asymptotic DA, $\phi_{as}(x) = 6 x (1-x)$ (dotted line),
and the AdS/CFT prediction of the asymptotic DA, 
$\phi_{AdS/CFT}(x) = \frac{8}{\pi}\sqrt{x(1-x)}$ (double-dot-dashed line).
While the quark DA for the HO potential parameter set 
lies between $\phi_{as}$ and $\phi_{AdS/CFT}$, that for 
the linear confining potential gets very close to $\phi_{AdS/CFT}$. 
In other words,
the central(end) point region is rather 
enhanced(suppressed) for the HO potential parameter set (1)  than the linear
potential paramter set (2). 
From our model calculation, we find that the shape of the quark DA becomes 
sharper and more suppressed at the endpoint region as the constituent
quark mass(gaussian parameter $\beta$) 
increases(decreases). 
As a sensitivity check, we note that the central point of the quark 
DA for the set (1) varies about 3.7$\%$(2.2$\%$) by changing 10$\%$ of 
the quark mass($\beta$ value).  This indicates that our results for the 
quark DA are quite stable for the variation of model parameters.

\begin{figure}[t]
\vspace{0.7cm}
\includegraphics[height=3in,width=3in]{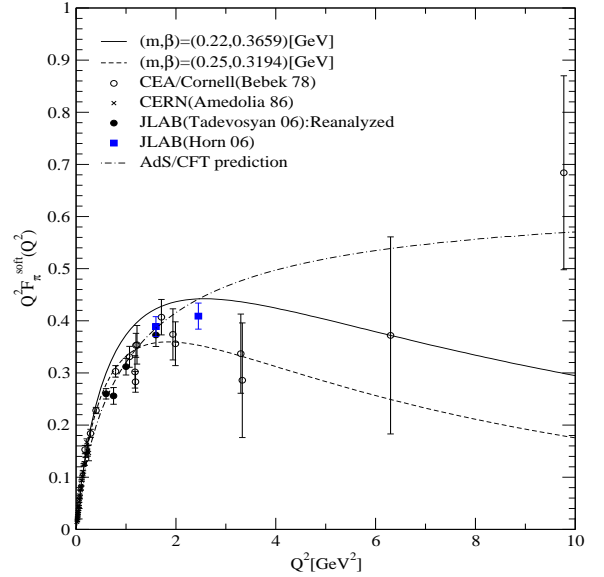}
\caption{Soft pion form factor $Q^2F_\pi(Q^2)$ in our linear(solid line) and
HO(dashed line) potential models compared with 
AdS/CFT prediction(dot-dashed line) as well as 
data~\cite{Bebek,Amen,Jlab061,Jlab062}.
Tadevosyan et al.~\cite{Jlab062} have reanalyzed 
previously published data by Volmer et al.~\cite{volmer} and obtained
the data points represented by the filled circles.}
\label{fig3}
\end{figure}
In Fig.~\ref{fig3}, we show our previously published result~\cite{CJ99} for
$Q^2F_\pi(Q^2)$ obtained from our linear(solid line) and HO(dashed line) 
confining potential models and compare both with the AdS/CFT 
prediction(dot-dashed line)\cite{GuyStan} and the 
data~\cite{Bebek,Amen,Jlab061,Jlab062}, which includes the most recent 
results from JLAB~\cite{Jlab061,Jlab062}. 
As expected, the gaussian fall-off at high $Q^2$ region provided by our 
previous LFQM results\cite{CJ99} with $m=0.22$ GeV (solid line) and
$m=0.25$ GeV (dashed line) cannot match with the power-law behavior
of the AdS/CFT prediction (dot-dashed line) although all the lines
(solid, dashed, dot-dashed) are comparable for $Q^2$ up to a few GeV$^2$.
While the recent
JLAB data from Horn et al.~\cite{Jlab061} extracted two new values at
$Q^2=1.60$ and 2.45 GeV$^2$, Tadevosyan et al.~\cite{Jlab062} 
have
reanalyzed the previously published values~\cite{volmer} of 
$Q^2=0.6-1.6$ GeV$^2$ and obtained rather lower values than those in 
Ref.~\cite{volmer}. 
We should note that our model parameters have been determined without 
the use of the pion form factor $F_\pi$ data but from the variational 
principle for the 
QCD-motivated effective Hamiltonian to fit the meson mass spectra.  
While our HO potential model(dashed line) provides a 
very good description of the reanalyzed values from
JLAB~\cite{Jlab062} up to $Q^2=1.60$ GeV$^2$, two newly extracted values
from~\cite{Jlab061}(filled squares) seem closer to our result from the 
linear potential model(solid line)
or the AdS/CFT prediction(dot-dashed line).
As shown in~\cite{Jlab061,Jlab062}, various model 
predictions such as Dyson-Schwinger~\cite{DS2}, QCD Sum Rules~\cite{NR}, 
and dispersion relation~\cite{Ge} including our linear potential model
slightly overestimate the pion form factor compare to
the new JLAB data~\cite{Jlab061,Jlab062}. 
Although it may be difficult to pin down which of our models
(linear or HO) is better to describe 
the current available data,  we at least note that the suppression
of the quark DA at the end points region corresponds to the 
suppression of the soft contribution to the pion form factor 
at low $Q^2$ region. This is rather similar to the case of the Sudakov
suppression enhancing the hard contribution to the form 
factor\cite{LS,Jacob,JK,ADT}.

\begin{figure}[t]
\vspace{0.7cm}
\includegraphics[height=3in,width=3in]{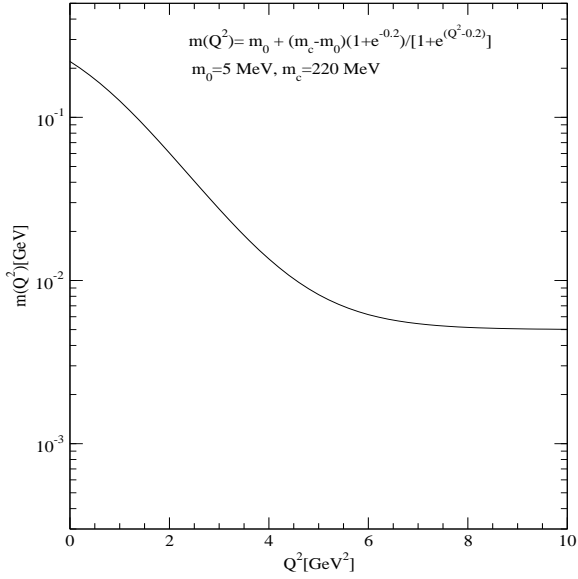}
\caption{Quark mass evolution $m(Q^2)$ in spacelike momentum region.
The $m_0$ and $m_c$ represent the current(at high $Q^2$) and 
constituent(at low $Q^2$) quark masses, respectively.}
\label{mass-evolution}
\end{figure}

\begin{figure}[t]
\vspace{0.7cm}
\includegraphics[height=3in,width=3in]{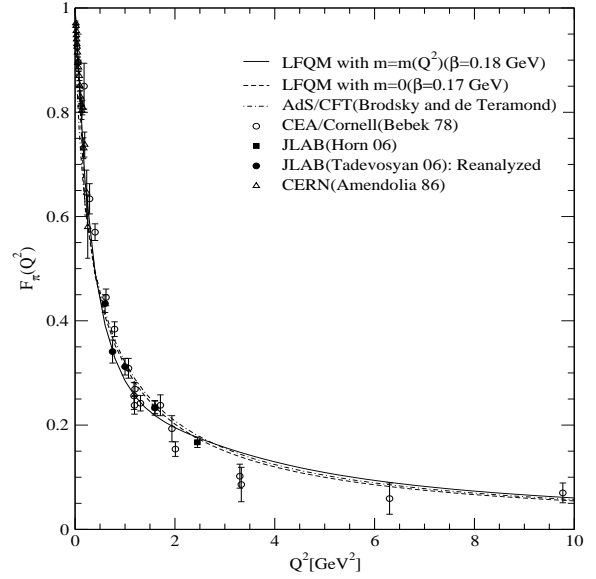}
\caption{$F_\pi(Q^2)$ with running mass $m(Q^2)$(solid line) and 
negilgible mass $m=0$(dashed line) in our LFQM (Eq.(\ref{soft_ff})) compared
to the AdS/CFT result~\cite{GuyStan}(dot-dashed line).}
\label{fig4}
\end{figure}

To investigate a consistency with the recent findings from the AdS/CFT 
correspondence~\cite{BT,GuyStan}, 
we consider a momentum-dependent quark mass for the
calculation of $F_\pi(Q^2)$ in LFQM. 
Since it is beyond the scope of our work to find an explicit form
of $m(Q^2)$, we take a simple parametrization similar to
what we have already used in our previous work\cite{KisslingerChoiJi}
as shown in Fig.~\ref{mass-evolution}. The details of discussion for
this type of parametrization and the consistency with the PCAC relation 
\cite{GOR} can be found in Ref.\cite{KisslingerChoiJi}. 
For comparison, we also simply take a
negligible current quark mass $m=0$ respecting the conformal symmetry
in Eq.~(\ref{soft_ff}) and show both results in Fig.~\ref{fig4}.

In Fig.~\ref{fig4}, the solid line is the result of Eq.(\ref{soft_ff}) 
using the effective dynamical quark mass $m(Q^2)$ drawn in 
Fig.\ref{mass-evolution},
which respects the conformal symmetry at high $Q^2$ limit.
In comparison, the dashed line is the result of Eq.(\ref{soft_ff}) 
taking $m=0$ and the dot-dashed line is the AdS/CFT prediction presented
in Ref.\cite{GuyStan}.
The corresponding values of the gaussian parameter $\beta=0.17\sim0.18$ GeV 
are not much different from the gaussian parameter $\kappa=0.4$ GeV used 
in a holographic AdS Gaussian-modified-metric model\cite{GuyStan}, 
e.g., the correspondence between $\beta$ and $\kappa$ can be easily seen 
from the gaussian dependence of our model wave function 
$\Psi(x,{\bf k}_\perp)\sim e^{-{\bf k}^2_\perp/2(2\beta)^2x(1-x)}$ in $m=0$ limit 
and the AdS Gaussian-modified-metric model wave function 
$\Psi(x,{\bf k}_\perp)\sim e^{-{\bf k}^2_\perp/2\kappa^2x(1-x)}$~\cite{CJ-DB}.
The decay constants obtained with these gaussian parameters are also
not much different from the experimental value $f_\pi = 92.4$ MeV. 
As shown in Fig.\ref{fig4}, our results respecting the conformal symmetry
are in excellent agreement with the AdS/CFT prediction\cite{GuyStan}.
This implies that, in the
massless quark case, our gaussian wave function with the Jacobi factor 
leads to
the scaling behavior $F_\pi(Q^2)\sim 1/Q^2$ as $Q^2\to\infty$,
being consistent with the scaling behavior obtained from AdS/CFT
correspondence~\cite{BT} at large $Q^2$. 
Although the power-law behavior $Q^n F_\pi(Q^2)\to const.$ for large $Q^2$ is
still obtained even without the Jacobi factor, we find 
$n=4$ rather than
$n=2$ if the Jacobi factor is not taken into account. 
The asymptotic behavior of $n=4$ for the meson form factor with
spin-1/2 quarks has recently been found using meromorphization for 
spin-1/2 quarks\cite{Ra_AdS}. Our work here indicates that the power
$n$, however, does depend on whether one takes into account the Jacobi 
factor or not in the LFQM calculation. Our result shows that $n=2$ can be
attained by taking into account the Jacobi factor in Eq.~(\ref{soft_ff}). 
This also distinguishes our analysis from Ref.~\cite{HW}. 
Other effects of the presence/absence
of the Jacobi factor to various static properties of mesons can be found in
Ref.~\cite{CJ_Jacob}.

\begin{figure}[t]
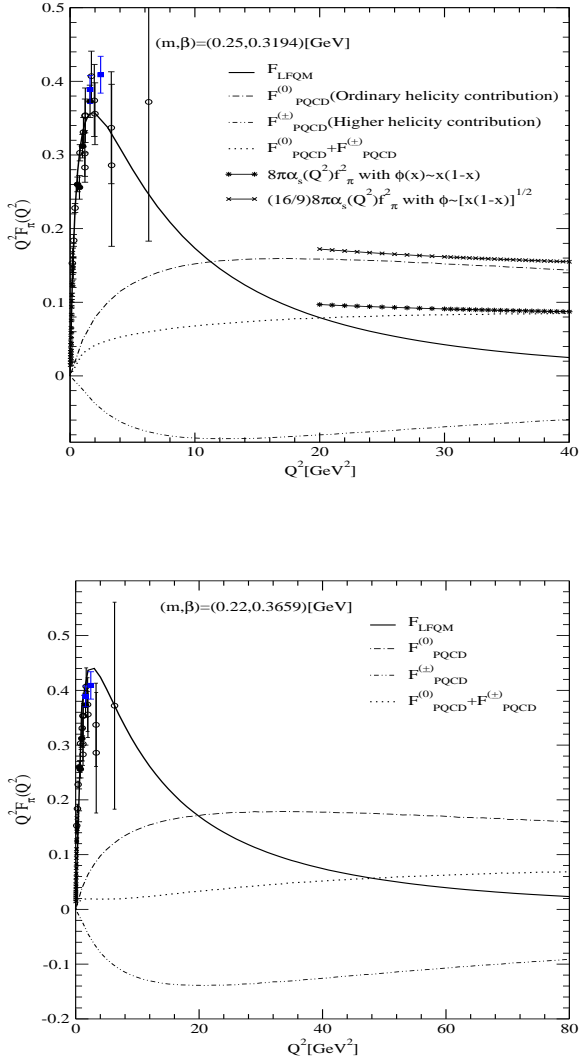

\vspace{0.7cm}
\includegraphics[height=2.5in,width=3in]{Fig6.eps}

\vspace{1.3cm}
\includegraphics[height=2.5in,width=3in]{Fig7.eps}
\caption{The hard(PQCD) and soft(LFQM) contributions to the pion form factor
$Q^2F_{\pi}(Q^2)$ using two different sets of model parameters,
(1) $(m, \beta)=(0.25, 0.3194)$ GeV[top], and
(2) $(m, \beta)=(0.22, 0.3659)$ GeV[bottom].}
\label{fig5}
\end{figure}
In Fig.~\ref{fig5}, 
we show the hard(PQCD) and soft(LFQM) contributions to the pion
form factor $Q^2F_{\pi}(Q^2)$ obtained from the HO[top] and  
linear[bottom] potential parameters, respectively. 
The solid and dotted lines represent the soft and 
hard(ordinary+higher helicities) contributions to the
pion form factor, respectively. We include each helicity component
contributions, i.e.  the ordinary(dot-dashed line) and 
higher(doubledot-dashed line) helicity contributions.
In the top panel, we also show the 
leading twist
PQCD predictions using both
$\phi_{as}$ and $\phi_{AdS/CFT}$ for the comparison.
As noted in Ref.\cite{GuyStan}, the broader shape of $\phi_{AdS/CFT}$ 
increases the magnitude of the leading twist PQCD prediction for the pion form 
factor by a factor of 16/9 compared to the prediction based on the asymptotic 
form.
The new experimental data
are taken from~\cite{Jlab061,Jlab062}.
In the AdS/CFT side, it is still an open question how
the AdS/CFT prediction of $Q^2 F_\pi (Q^2)$ shown in Fig.\ref{fig3}
can approach asymptotically to the PQCD prediction using $\phi_{AdS/CFT}$
shown in Fig.\ref{fig5}~\cite{CJ-DB}.
Possibly, the higher order PQCD corrections may reconcile the two approaches
and in addition one may need to evolve the AdS/CFT DA to the scale of the 
gluon virtuality. Since the loop contributions are not included to the
the present approximation of the AdS/CFT correspondence, quantum corrections
are to be incorporated further~\cite{CJ-DB}. It is not yet clear how to do this
in practice, although there is much interest in the theme~\cite{Csaki-Reece}.
In the QCD side, however, 
there are a few  
things to note from Fig.~\ref{fig5} regarding on the soft/hard 
contribution: 
(a) The soft contribution for the linear potential parameters is larger
than that for the HO potential ones. 
(b) The hard contribution for the linear potential parameters is smaller
than that for the HO potential ones.
(c) The higher helicity components suppress the contributions from the
ordinary helicity components. 
From (a) and (b), we find that the soft and hard contributions are correlated 
to each other, i.e. as the endpoint region for the quark DA is more suppressed,  
the soft(hard) contribution to the pion form factor does get 
less(more) enhancement. This finding is rather similar to the previous
findings from the Sudakov suppression of the soft 
contribution\cite{LS,Jacob,JK,ADT}.
From (c), we also see that the intrinsic transverse momentum effect included
in this work is still effective even 
at $Q^2 \approx 80$GeV$^2$ range.

\section{Summary and Conclusion}
Due to the ongoing debate on the PQCD applicability in exclusive processes, 
we studied the soft and hard contributions to the pion form factor utilizing 
both LFQM and PQCD approaches.

We discussed a constraint of conformal symmetry in the analysis of the pion 
form factor.  The usual power-law behavior of the pion form factor obtained 
in the perturbative QCD analysis can also be attained by taking negligible 
quark masses in the nonperturbative quark model analysis, confirming the 
recent AdS/CFT correspondence. Inclusion of the Jacobi factor with the 
gaussian radial wave function in the LFQM analysis is essential to 
attain the power-law behavior $F_\pi(Q^2)\sim 1/Q^2$
in the massless quark limit. 
We find that the correlation between the shape of nonperturbative quark 
distribution amplitude and the amount of soft and hard contributions to the 
form factor, i.e. the suppresion(enhancement) of the endpoint region for the 
quark distribution amplitude corresponds to the suppression of the soft(hard) 
contribution. The fact that the higher helicity components suppress the 
contributions from the ordinary helicity components are also confirmed in 
this work.

The conformal symmetry as well as the correlation between 
soft and hard contributions may provide a 
useful constraint on the model building for the form factor analysis.

\acknowledgements
We thank Stanley J. Brodsky and Guy F. de Teramond for many useful 
discussions.  
This work was supported by a grant from the U.S. Department of 
Energy(No. DE-FG02-96ER40947).  H.-M. Choi was supported in part by Korea 
Research Foundation under the contract KRF-2005-070-C00039. The National 
Energy Research Scientific Center is also acknowledged for the grant of 
supercomputing time. We also acknowledge the Asia Pacific Center for 
Theoretical Physics(APCTP) for the support on the APCTP Focus Program during
which this work was completed.


\begin{thebibliography}{99}
\bibitem{BL} G. P. Lepage and S. J. Brodsky, \Journal{\PRD}{22}{2157}{1980}.
\bibitem{IL} N. Isgur and C. H. Llewellyn Smith, 
\Journal{\PRL}{52}{1080}{1984}; \Journal{\NPB}{317}{526}{1989}.
\bibitem{Ra} A. V. Radyushkin, \Journal{\NPA}{532}{141}{1991}.
\bibitem{LS} H. N. Li and G. Sterman, \Journal{\NPB}{381}{129}{1992};
H. N. Li, \Journal{\PRD}{48}{4243}{1993}.
\bibitem{Jacob} O.C. Jacob and L.S. Kisslinger,\Journal{\PRL}{56}{225}{1986};
\Journal{\PLB}{243}{323}{1990}.
\bibitem{JK} R. Jakob and P. Kroll, \Journal{\PLB}{315}{463}{1993}.
\bibitem{ADT} V. Anikin, A.-E. Dorokhov and L. Tomio,
\Journal{\PLB}{475}{361}{2000}.
\bibitem{SSK} N.G. Stefanis, W. Schroers, and H.-C. Kim, 
\Journal{\EPJC}{18}{137}{2000}.
\bibitem{HW} T. Huang, X.-G. Wu, and X.-H. Wu, \Journal{\PRD}{70}{053007}{2004}.
\bibitem{JPS} C.-R. Ji, A. Pang, and A. Szcepaniak,
\Journal{\PRD}{52}{4038}{1995}.
\bibitem{SJR} A. Szcepaniak, C.-R. Ji, and A. Radyushkin,
\Journal{\PRD}{57}{2813}{1998}.
\bibitem{BT} S. J. Brodsky and Guy F. de Teramond, 
\Journal{\PLB}{582}{211}{2004}; \Journal{\PRL}{96}{201601}{2006}.
\bibitem{Ra_AdS} A. V. Radyushkin, hep-ph/0605116.
\bibitem{CJ99} H.-M. Choi and C.-R. Ji, \Journal{\PRD}{59}{074015}{1999}.
\bibitem{Mel} H. J. Melosh, \Journal{\PRD}{9}{1095}{1974}.
\bibitem{PJ} P. J. O'Donnell, Q.P. Xu, and H.K.K. Tung, 
\Journal{\PRD}{52}{3966}{1995}; W. Jaus, \Journal{\PRD}{53}{1349}{1996};
I.L. Grach, I.M. Narodetskii, and S. Simula, \Journal{\PLB}{385}{317}{1996};
H.-Y. Cheng, C.-Y. Cheng, and C.-W. Hwang, \Journal{\PRD}{55}{1559}{1997}.
\bibitem{Wil} 
J.I. Skullerud and A.G. Williams, \Journal{\PRD}{63}{054508}{2001};
J.B. Zhang {\em et al.}, \Journal{\PRD}{70}{034505}{2004}.
\bibitem{DS1} P. Maris and C.D. Roberts, \Journal{\PRC}{58}{3659}{1998}.
\bibitem{DS2} P. Maris and P.C. Tandy, \Journal{\PRC}{62}{204}{2000}.
\bibitem{DD} H.-M. Choi and C.-R. Ji, \Journal{\PRD}{73}{114020}{2006}.
\bibitem{GuyStan} G. F. de Teramond, Talk entitled ``Mapping AdS/CFT Results for
Holographic QCD to the Light Front" presented in the Workshop LC2006 (Light-Cone 
QCD and Nonperturbative Hadron Physics), Minneapolis, May 15-19, 2006; 
S. J. Brodsky, hep-ph/0608005 and hep-ph/0610115.
\bibitem{Bebek} C. J. Bebek, {\em et al.}, \Journal{\PRD}{17}{1693}{1978}.
\bibitem{Amen} S.R. Amendolia, {\em et al.}, \Journal{\NPB}{277}{168}{1986}.
\bibitem{Jlab061} T. Horn, {\em et al.}, nucl-ex/0607007.
\bibitem{Jlab062} V. Tadevosyan, {\em et al.}, nucl-ex/0607007.
\bibitem{volmer} J. Volmer {\em et al.}, \Journal{\PRL}{86}{1713}{2001}.
\bibitem{NR} V. A. Nesterenko and A.V. Radyushkin, 
\Journal{\PLB}{115}{410}{1982}.
\bibitem{Ge} B.V. Geshkenbein, \Journal{\PRD}{61}{033009}{2000}.
\bibitem{KisslingerChoiJi} L.S. Kisslinger, H.-M. Choi and C.-R. Ji,
\bibitem{GOR} M.Gell-Mann, R.J.Oakes and B.Renner, Phys. rev. {\bf 175},
2195 (1968).
\Journal{\PRD}{63}{113005}{2001}. 
\bibitem{CJ-DB} Private communications with S. J. Brodsky and G. F.
de Teramond.
\bibitem{CJ_Jacob} H.-M. Choi and C.-R. Ji, \Journal{\PRD}{56}{6010}{1997}.
\bibitem{Csaki-Reece} C. Csaki and M. Reece, hep-ph/0608266.
\end{thebibliography}
\end{document}